\documentclass[a4paper,fleqn]{cas-dc}

\usepackage[version=3]{mhchem} 
\usepackage{color}
\usepackage{graphicx}
\usepackage{lineno,hyperref}
\usepackage[numbers]{natbib}

\begin{document}
\let\WriteBookmarks\relax
\def\floatpagepagefraction{1}
\def\textpagefraction{.001}
\shorttitle{Infrared FEL irradiation of CO$_{2}$ ice}
\shortauthors{S Ioppolo et~al.}

\title [mode = title]{Infrared free-electron laser irradiation of carbon dioxide ice}



\author[1]{Sergio Ioppolo}[orcid=0000-0002-2271-1781]
\cormark[1]
\ead{s.ioppolo@qmul.ac.uk}
\cortext[cor1]{Corresponding author}

\author[2,3]{Jennifer A. Noble}[orcid=0000-0003-4985-8254]
\author[1]{Alejandra Traspas~Mui\~{n}a}[orcid=0000-0002-4304-2628]
\author[4]{Herma M. Cuppen}[orcid=0000-0003-4397-0739]
\author[2]{St\'{e}phane Coussan}[orcid=0000-0002-0275-7272]
\author[5]{Britta Redlich}[orcid=0000-0002-7770-9531]

\address[1]{School of Electronic Engineering and Computer Science, Queen Mary University of London, London E1 4NS, UK}
\address[2]{CNRS, Aix-Marseille Univ, PIIM, Marseille 13397, France}
\address[3]{School of Physical Sciences, University of Kent, Canterbury CT2 7NH, UK}
\address[4]{Radboud University, Institute for Molecules and Materials, Nijmegen 6525 AJ, The Netherlands}
\address[5]{FELIX Laboratory, Radboud University, Nijmegen 6525 ED, The Netherlands}

\begin{abstract}
Interstellar ice grains are believed to play a key role in the formation of many of the simple and complex organic species detected in space. However, many fundamental questions on the physicochemical processes linked to the formation and survival of species in ice grains remain unanswered. Field work at large-scale facilities such as free-electron lasers (FELs) can aid the investigation of the composition and morphology of ice grains by providing novel tools to the laboratory astrophysics community. We combined the high tunability, wide infrared spectral range and intensity of the FEL beam line FELIX-2 at the HFML-FELIX Laboratory in the Netherlands with the characteristics of the ultrahigh vacuum LISA end station to perform wavelength-dependent mid-IR irradiation experiments of space-relevant pure carbon dioxide (CO$_{2}$) ice at 20~K. We used the intense monochromatic radiation of FELIX to inject vibrational energy at selected frequencies into the CO$_{2}$ ice to study ice restructuring effects \emph{in~situ} by Fourier Transform Reflection-Absorption Infrared (FT-RAIR) spectroscopy. This work improves our understanding of how vibrational energy introduced by external triggers such as photons, electrons, cosmic rays, and thermal heating coming from a nascent protostar or field stars is dissipated in an interstellar icy dust grain in space. Moreover, it adds to the current literature debate concerning the amorphous and polycrystalline structure of CO$_{2}$ ice observed upon deposition at low temperatures, showing that, under our experimental conditions, CO$_{2}$ ice presents amorphous characteristics when deposited at 20~K and is unambiguously crystalline if deposited at 75~K.
\end{abstract}




\maketitle

\section{Introduction}
To date, more than 240 different species have been identified in space in the gas phase \cite{McGuire:2021}. Since gas-phase reaction rates are often found to be too low to account for the observed abundances of species in space, the icy mantles of interstellar dust grains are believed to give rise to a rich prebiotic chemistry, which feeds complex organic molecules (COMs) to star-forming regions and ultimately to planetary systems \cite{Herbst:2009, Oberg:2021, Ioppolo:2021}. Current infrared (IR) observations show that laboratory spectral data on peak position and feature width can be used to extract information on layering and mixing of interstellar ices \cite{Penteado:2015, Boogert:2015}. The NASA James Webb Space Telescope (JWST) mission will enable the observation of interstellar ices over a wide wavelength range (0.6-28.3 $\mu$m) toward a large variety of sources. Its high spectral resolution will allow the observation of spectral characteristics depending on different ice morphologies, thermal histories, and mixing environments in star-forming regions, beyond simply measuring feature width and peak position. However, the large amount of new observational data that will be provided by JWST can only be correctly interpreted if a parallel effort is made in laboratories to develop new experimental methodologies, for instance, at free-electron laser (FEL) large-scale facilities, to provide unprecedented spectroscopic details on morphologies and dynamics within interstellar ice analogs.

Believed to be largely amorphous in nature, the structure of interstellar ices is still under investigation. In space, dust grains covered in icy material are continuously exposed to ‘energetic’ particles such as cosmic rays, electrons, and UV photons that release excess energy into the ice target causing physicochemical changes within the grain mantle. Exothermic surface chemical reactions induced by gas-phase species landing on the grain surface can also affect the structure and dynamics within interstellar ices. Understanding how exothermic chemical reactions deposit large amounts of energy into icy mantles covering dust grains containing some of the molecular building blocks for life, could help explain the formation of prebiotic compounds in space. Particularly, surface chemical reactions can be broken down into various elementary processes, namely, adsorption, diffusion, dissociation, product formation, energy dissipation, and desorption. The mechanisms underlying energy dissipation at the atomic scale thus decide the fate of possible subsequent processes.

To date, little is known about how the energy released into an icy grain after a surface reaction or cosmic ray interaction dissipates within the grain \cite{Ivlev:2015}. Direct experimental quantification of energy dissipation by time-resolved techniques with concurrent atomic resolution is still an ongoing challenge \cite{Dell'Angela:2013, Ostrom:2015}. Moreover, the effect of low energy photon irradiation, \emph{i.e.}, in the infrared regime, on -- for instance -- the hydrogen bonding network of water ice has been scarcely addressed. Only a couple of IR irradiation studies have been performed on crystalline water ice at elevated temperatures, the main purpose of which was to investigate desorption efficiencies and products \cite{Krasnopoler:1998, Focsa:2003}. In terms of porous amorphous solid water (pASW), some of us have previously studied the pASW surface via  the  ‘dangling’  OH  stretching  modes \cite{Noble:2014a, Noble:2014b, Coussan:2015}, revealing an irreversible restructuring of these surface modes upon selective IR irradiation by means of a table-top pulsed optical parametric oscillator (OPO) laser. Our recent combined laboratory and theoretical work on selective FEL irradiation of pASW showed that ices can absorb vibrational energy in the broad IR (4000-400 cm$^{-1}$, 2.5-25 $\mu$m) and terahertz (THz, 0.1-10 THz, 30-3000 $\mu$m) spectral ranges, causing local heating within the ice structure and at the surface that leads to a reorganization of the ice toward a more ordered material \cite{Noble:2020}.

Selective IR/THz experiments at FEL facilities are needed to better understand the nature of IR/THz modes in solids as well as to time-resolve dynamics and energy relaxation processes within condensed matter. Particularly important to the fields of astrochemistry, astrobiology, surface science, and catalysis are physicochemical studies of diffusion, segregation, reaction, and ultimately desorption of molecules at the surface and in the bulk of space-relevant ice layers. Here, we present the first laboratory selective FEL irradiation study of pure solid amorphous and crystalline carbon dioxide (CO$_{2}$) ices at 20~K. Based on our previous work on ASW, cubic ice (Ic), and hexagonal ice (Ih), we selected CO$_{2}$ as our target for the following reasons. ASW is the most abundant form of ice in the Universe and provides the surface upon, and matrix within which, volatile species react to form prebiotic molecules, as well as absorbing some of the excess reaction energy \cite{Fraser:2001, Noble:2014b}. After H$_{2}$O, CO$_{2}$ is one of the most abundant solid-phase species in the interstellar medium (ISM), with abundances of between 20 and 30\% that of H$_{2}$O ice, depending on the object observed \cite{Boogert:2015}. In dense molecular clouds, CO$_{2}$ is formed in the solid phase primarily through the CO+OH surface reaction \cite{Goumans:2008, Oba:2010, Ioppolo:2011, Noble:2011}. Therefore, CO$_{2}$ tends to reside mostly in a H$_{2}$O-rich layer, where OH radicals are more abundant and available for reaction with CO molecules. Additionally, CO$_{2}$ and more complex species form in mixed ices that undergo ‘energetic’ processing such as thermal heating, photon, electron, and ion irradiation that dramatically affect the bulk and surface structure of the ice \cite{Moore:1991, Mennella:2004, Mennella:2006, Loeffler:2005, Gomis:2005, Ioppolo:2009, Ioppolo:2013, Suhasaria:2017, Arumainayagam:2019}. Therefore, understanding changes induced upon chemical or physical perturbation in CO$_{2}$ ices is pivotal to unraveling molecular evolution in space.

Finally, there is an ongoing debate in the literature on the structure of deposited CO$_{2}$ ice under laboratory conditions and its corresponding spectroscopic fingerprint assignments that can help interpreting present and future observations of CO$_{2}$ ice in space \cite{Escribano:2013, Allodi:2014, Gerakines:2015, McGuire:2016, Baratta:2017, Tsuge:2020, Kouchi:2021}. This work adds to the scientific discussion on the topic by providing new independent evidence of the morphology of CO$_{2}$ ice deposited at different temperatures with implications for the ISM and the Solar System.

\section{Experimental Methods}
Experiments are carried out using the laboratory ice surface astrophysics (LISA) ultrahigh vacuum (UHV) end station at the HFML-FELIX Laboratory, Radboud University, The Netherlands. Details of the experimental setup are given in the Supporting Material. Here we discuss the experimental methodology used during a standard experiment. Briefly, pure CO$_{2}$ ices are grown on the gold substrate of the main chamber at temperatures of 20 and 75~K. Carbon dioxide ($99.995\%$ CO$_{2}$, CANgas) is used as received from Sigma-Aldrich. Firstly, a few mbar of CO$_{2}$ gas are introduced in the dosing line, while monitoring the pressure mass-independently. Then, gases are admitted into the main chamber by a manual all-metal needle valve. During deposition, the pressure in the main chamber is measured by means of the cold cathode gauge and the QMS and it is kept constant at 1$\times$10$^{-6}$~mbar by adjusting the aperture of the all-metal inlet valve. QMS calibration experiments are performed at room temperature to find a correlation between residual mass signals detected with the cryostat on and off.

\begin{figure}
	\centering
	\includegraphics[width=\hsize]{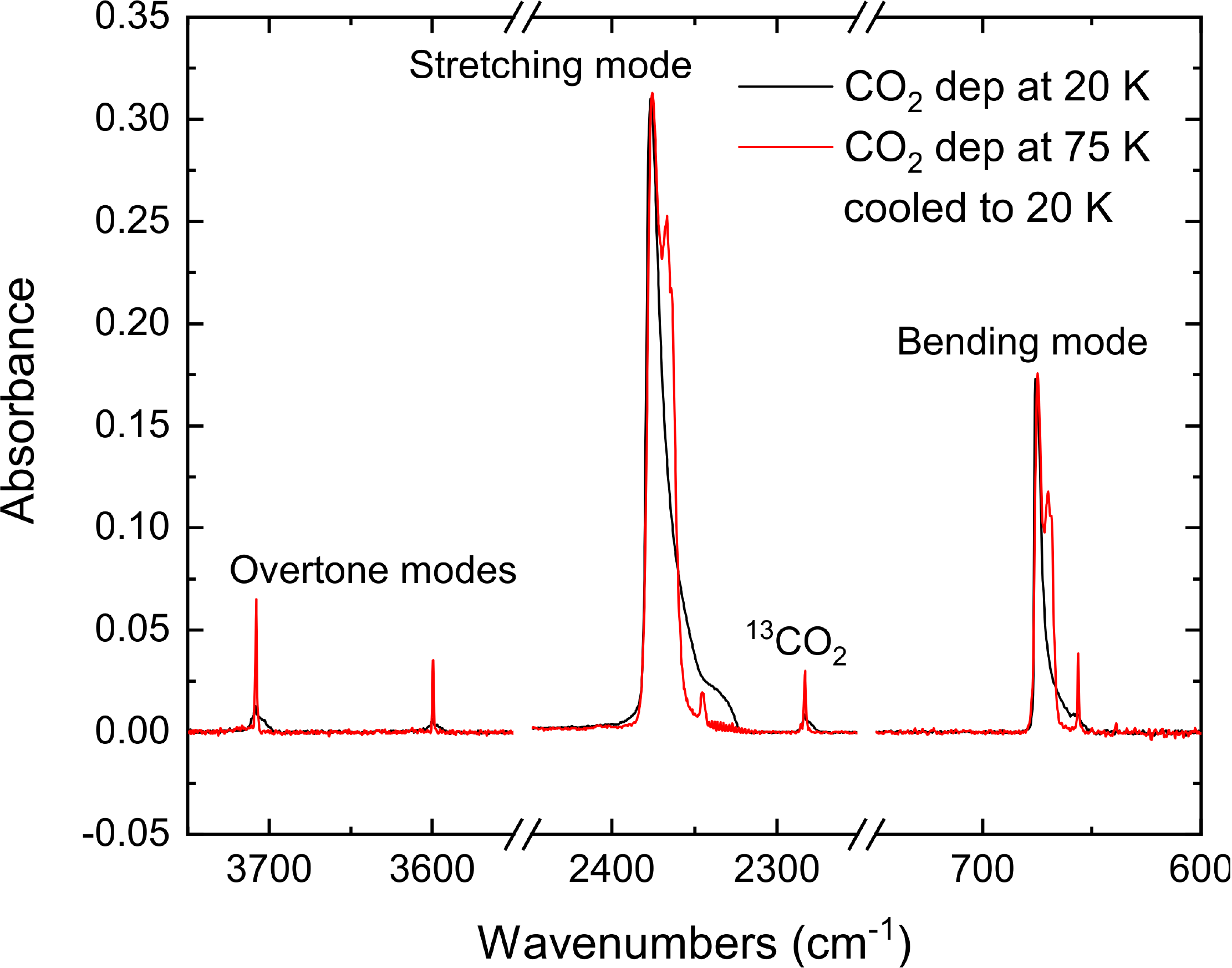}
	\caption{FT-RAIR spectra of pure CO$_{2}$ deposited at 20~K (black solid line) and at 75~K and then cooled to 20~K (red solid line) marked with absorption modes in the MIR. Solid $^{13}$CO$_{2}$ is also visible.
	}
	\label{Fig2}
\end{figure}

At the beginning of a standard experiment, the cryostat is turned on, the MCT detector is filled with liquid nitrogen, and the nitrogen purge is activated. Once the base temperature of the substrate is reached in a matter of a few hours, the gold substrate is flash heated to 200~K to further clean its surface prior to gas deposition. Then, upon reaching the selected substrate temperature (20 or 75~K), a background FTIR spectrum comprising 512 co-added scans with a resolution of 0.5~cm$^{-1}$ is acquired in the 4000-600~cm$^{-1}$ range to remove any signal along the IR beam path that does not arise from the ice sample. During deposition, FTIR spectra with 8 co-added scans are acquired every 20 seconds to monitor the ice growth, while residual gas is measured by the QMS. At the end of all depositions, an FTIR spectrum with 256 co-added scans is acquired. If the deposition temperature is not 20~K, the ice is then cooled and kept at this temperature and another FTIR spectrum with 256 co-added scans is acquired prior to any exposure to FEL radiation. All spectra are acquired at a resolution of 0.5~cm$^{-1}$. Figure~\ref{Fig2} shows the FT-RAIR spectra of pure CO$_{2}$ deposited at 20~K and at 75~K and then cooled to 20~K. The figure highlights the overtone ($\nu_1 + \nu_3$ at 3708~cm$^{-1}$ and $2\nu_2 + \nu_3$ at 3599~cm$^{-1}$), stretching ($\nu_3$ at 2376~cm$^{-1}$), and bending ($\nu_2$ at 675~cm$^{-1}$) modes of pure $^{12}$CO$_{2}$ as well as the natural presence of $^{13}$CO$_{2}$ ($\nu_3$ at 2283~cm$^{-1}$) in the deposited sample. Differences in spectral profiles and absorption peak positions compared to some literature values are due to the fact that here spectra are acquired by means of RAIR spectroscopy as opposed to transmission spectroscopy \cite{Gerakines:2015, Baratta:2017}.

\begin{table}
\centering
        \caption{Experimental details of CO$_{2}$ ice exposed to FEL radiation.}
        \label{Table1}
        \begin{tabular}{llll}
                \hline
        Molecule & Deposition T & FEL irradiation & TPD \\
                 & K & $\mu$m & K~min$^{-1}$\\
        \hline
        CO$_{2}$ & 20  & 3.2, 4.19, 4.21,  & 2.5\\
                 &     & 4.22, 4.28, 14.76,&    \\
                 &     & 15.18             &    \\
        CO$_{2}$ & 20  & ---               & 2.5\\
        CO$_{2}$ & 75  & 3.2, 4.2, 4.22,   & ---\\
                 &     & 4.25, 14.8        &    \\
                \hline
        \end{tabular}
\end{table}

The molecular column density $C$ (molecules cm$^{-2}$) of the deposited CO$_{2}$ ices at both 20 and 75~K is calculated by integrating a modified Beer-Lambert Equation (Eq. 1) to account for the RAIR mode configuration over the $\nu_3$ absorbance band of CO$_{2}$:

\begin{equation}
 C = \ln(10) \frac{\sin(\theta)}{2 A'(\nu)} \int{\tau(\nu) d\nu},
\end{equation}
where $A'(\nu)$ is the integrated band strength, $\tau(\nu)$ is the optical depth (cm$^{-1}$), and $\theta$ is the FTIR incident angle of 13$^\circ$. Based on the work of Gerakines and Hudson~\cite{Gerakines:2015}, we chose $A'(\nu_3)~=~1.18\times10^{-16}$~cm~molecule$^{-1}$ and $A'(\nu_3)~=~7.6\times10^{-17}$~cm~molecule$^{-1}$ as the integrated band strengths for pure CO$_{2}$ deposited at 20~K and at 75~K and then cooled to 20~K, respectively \cite{Gerakines:2015, Yamada:1964, Gerakines:1995}. In our experiments, the column densities of CO$_{2}$ deposited at 20 and at 75~K are 11$\times10^{15}$ and 17$\times10^{15}$~molecule~cm$^{-2}$, respectively. Apart from the deposition temperature, all other parameters are kept the same during deposition, \emph{e.g.}, deposition time ($\sim$190~s) and residual pressure in the main chamber ($1\times10^{-6}$~mbar). Once calculated, the column densities are used to determine the ice thicknesses in nm as per Eq. 2:

\begin{equation}
 d = \frac{C Z}{\rho N_A} \times 10^7,
\end{equation}
where $Z$ is the molecular mass of the CO$_{2}$ molecule (g~mol$^{-1}$), $\rho$ is the density of the ice, which we have taken to be 1.2~g~cm$^{-3}$ and 1.67~g~cm$^{-3}$ for CO$_{2}$ deposited at 20~K and at 75~K, respectively \cite{Loeffler:2016}, and $N_A$ is the Avogadro constant ($6.02\times10^{23}$~molecule~mol$^{-1}$). Hence, we estimate pure CO$_{2}$ deposited at 20 and 75~K to be both $\sim$7~nm thick with a deposition rate of $\sim$0.037~nm~s$^{-1}$.

\begin{figure*}
	\centering
	\includegraphics[width=\hsize]{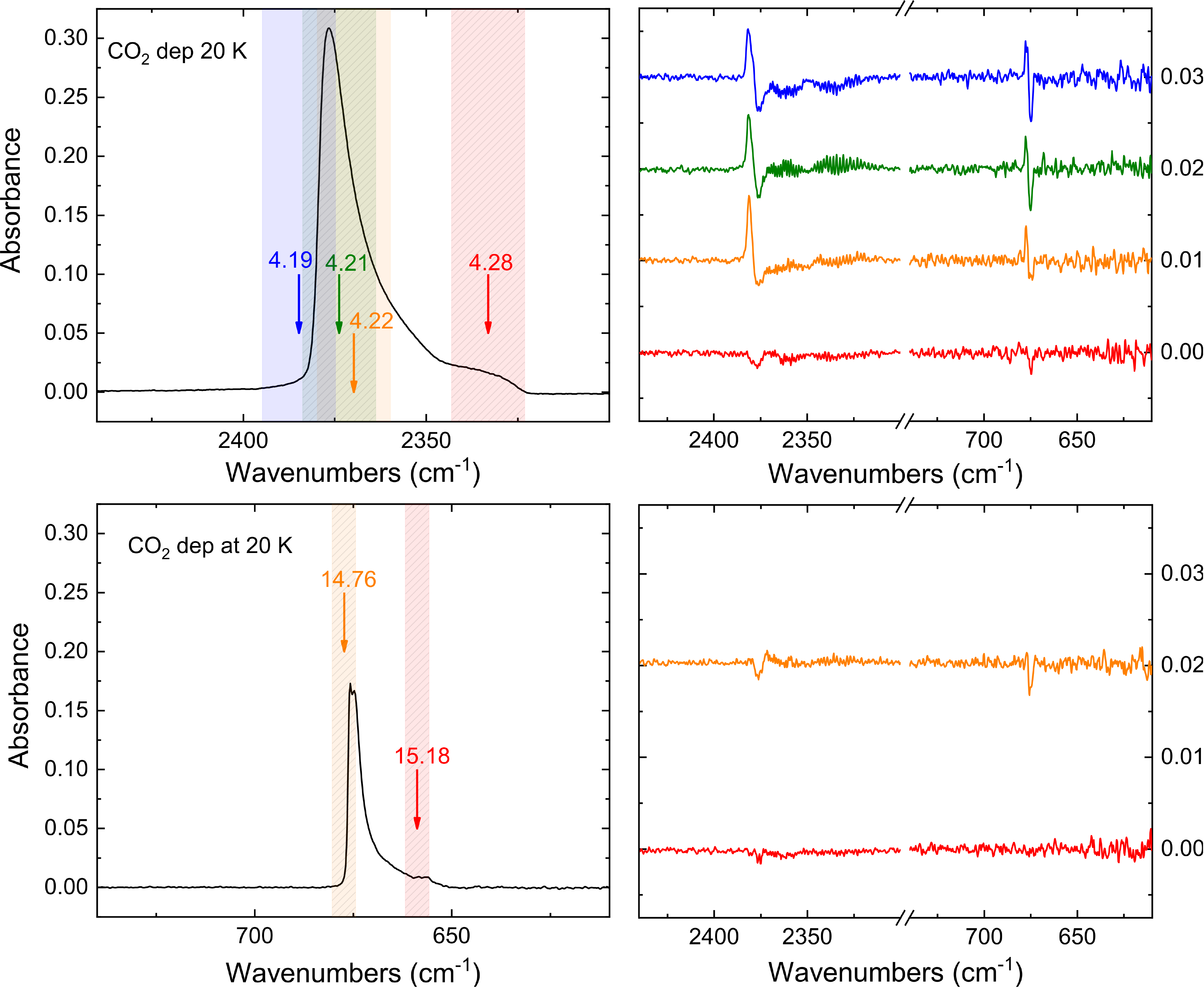}
	\caption{FEL irradiation of pure amorphous CO$_{2}$ ice at 20~K. Left panels show an FT-RAIR spectrum of pure CO$_{2}$ ice after deposition in the $\nu_3$ and $\nu_2$ spectral ranges (upper and lower panels, respectively). Colored arrows and strips indicate the frequencies and FWHM of FELIX-2 used during irradiation, respectively. Right panels show FT-RAIR difference spectra before and after FEL exposure. Spectra and FEL irradiation frequencies are color-coded. Difference spectra are offset for clarity.}
	\label{Fig3}
\end{figure*}

Ices are irradiated at 20~K using the FELIX-2 infrared FEL source with macropulses of 5 Hz repetition rate and at frequencies in the MIR range (3.2-15.2~$\mu$m).  All irradiations of pure CO$_{2}$ ice deposited at 20~K and at 75~K are carried out for five minutes to ensure the complete saturation of any structural changes in the ice. The spectral FWHM of the FELIX beam is on the order of 0.8~\%~$\delta\lambda/\lambda$ for all wavelengths. Unless otherwise specified, the laser macropulse energy at longer wavelengths ($\sim$130~mJ in the 15~$\mu$m wavelength region) is attenuated to $\sim$20~mJ to match that in the 3-4~$\mu$m region. Thus, at all wavelengths investigated, the average power is $\sim$0.03~W, with a laser fluence at the sample of approximately $\sim$0.2~J/cm$^2$. Table~\ref{Table1} shows the main experiments and their FEL irradiation details.

FTIR spectra are acquired before and after irradiation, from which difference spectra are obtained to highlight changes in the IR band profiles. All data analysis and spectral manipulation is performed using \emph{in~house} python scripts. Unless specified, FEL irradiations are performed on unirradiated ice. During FEL irradiation, the QMS is used to monitor any possible desorption of species from the ice. Upon completion of FEL irradiations, a TPD experiment is usually performed by means of the QMS as well as the FTIR with spectra of 128 co-added scans collected every 120 seconds. Since a heating ramp of 2.5~K~min$^{-1}$ is used, each FTIR spectrum is acquired every 5~K from 20 to 200~K. A TPD control experiment of a CO$_{2}$ ice deposited at 20~K and not exposed to FEL irradiation is performed for comparison and interpretation of the FEL exposure experiments.

\section{Results and discussion}

\subsection{FEL irradiation of CO$_{2}$ ice deposited at 20~K}
The left panels of Figure~\ref{Fig3} show an FT-RAIR spectrum of pure CO$_{2}$ ice deposited at 20~K in the $\nu_3$ and $\nu_2$ spectral ranges (upper and lower panels, respectively). Both $\nu_3$ and $\nu_2$ modes exhibit tranverse optical (TO) and longitudinal optical (LO) phonons in which the normal vibrations propagate through the ice lattice perpendicular and parallel to the direction of the IR ﬁeld, respectively. In our FT-RAIR spectra, the LO phonons at 4.2 and 14.8~$\mu$m are stronger than the TO phonons at 4.27 and 15.2~$\mu$m for the $\nu_3$ and $\nu_2$ modes, respectively \cite{Escribano:2013, Baratta:1998, Cooke:2016}. Colored arrows indicate the frequencies at which the ice has been exposed to FELIX-2 radiation for five minutes of continuous exposure at a time and always on fresh, unirradiated ice spots. Here and throughout the rest of the paper, we will use the same color scheme.  Vertical color-coded strips indicate the spectral FWHM of the FELIX beam at each selected wavelength. Around the peak of the $\nu_3$ mode, the spectral FWHM of the FELIX beam is such that at all three selected FEL irradiation frequencies the three irradiations overlap with one other. This can explain the similar effects seen in the ice after FEL exposure for the three irradiations at 4.19, 4.21, and 4.22~$\mu$m by comparing FT-RAIR spectra before and after FEL exposure (upper-right panel of Fig.~\ref{Fig3}). Such difference spectra reveal that the ice has undergone a degree of restructuring as well as perhaps some desorption. The latter is highlighted by the fact that, upon exposure to FEL radiation, both $\nu_3$ and $\nu_2$ modes decreased in intensity, as shown by the negative peaks in the upper three difference spectra of the upper-right panel of Fig.~\ref{Fig3}. Moreover, the $\nu_3$ and $\nu_2$ band shapes present modified profiles suggesting restructuring of the ice, as shown by the positive peaks in the three difference spectra. More pronounced negative peaks are observed for irradiations around the LO phonon mode peak of the $\nu_3$ band. The fourth FEL irradiation at 4.28 $\mu$m on the red wing (\emph{i.e.}, TO phonon) of the CO$_{2}$ stretching mode presents only weaker negative peaks. FEL irradiation can depend on the absorption cross section of the $\nu_3$ band, showing weaker effects at 4.28 $\mu$m. Moreover, a larger restructuring effect seen after FEL irradiations of the LO versus the TO phonon modes of the $\nu_3$ band can be explained by the fact that the FELIX-2 optical configuration used for the experiments presented here is such that the FEL radiation is p-polarized, \emph{i.e.} with the electric vector parallel to the incidence plane and in resonance with LO phonon modes.

FEL irradiations at the CO$_{2}$ bending mode (14.76 and 15.18 $\mu$m, lower-left panel of Fig.~\ref{Fig3}) provoke spectral changes that are comparable to those seen upon irradiating $\nu_3$ at its red wing, namely, negative peaks. Differences in spectral changes upon FEL radiation of the $\nu_2$ and $\nu_3$ bands suggest that on-resonance injection of vibrational energy into the CO$_{2}$ ice is a frequency dependent process. As in the case of the $\nu_3$ band, the effect is relatively stronger for FEL irradiation at the LO phonon mode peak of the $\nu_2$ mode and weaker at its red wing (\emph{i.e.}, TO phonon mode). Much like in the case of pASW, FEL irradiations performed at frequencies where the ice molecules do not present any vibrational absorption do not cause any change in the ice structure nor its thickness \cite{Noble:2020}. Hence, the difference spectra acquired before and after FEL exposure at non-resonant frequencies appear flat (spectra not shown in Fig.~\ref{Fig3}). This is a further confirmation that it is the ice that absorbs vibrational energy and not the gold substrate underneath.

\begin{figure*}
	\centering
	\includegraphics[width=\hsize]{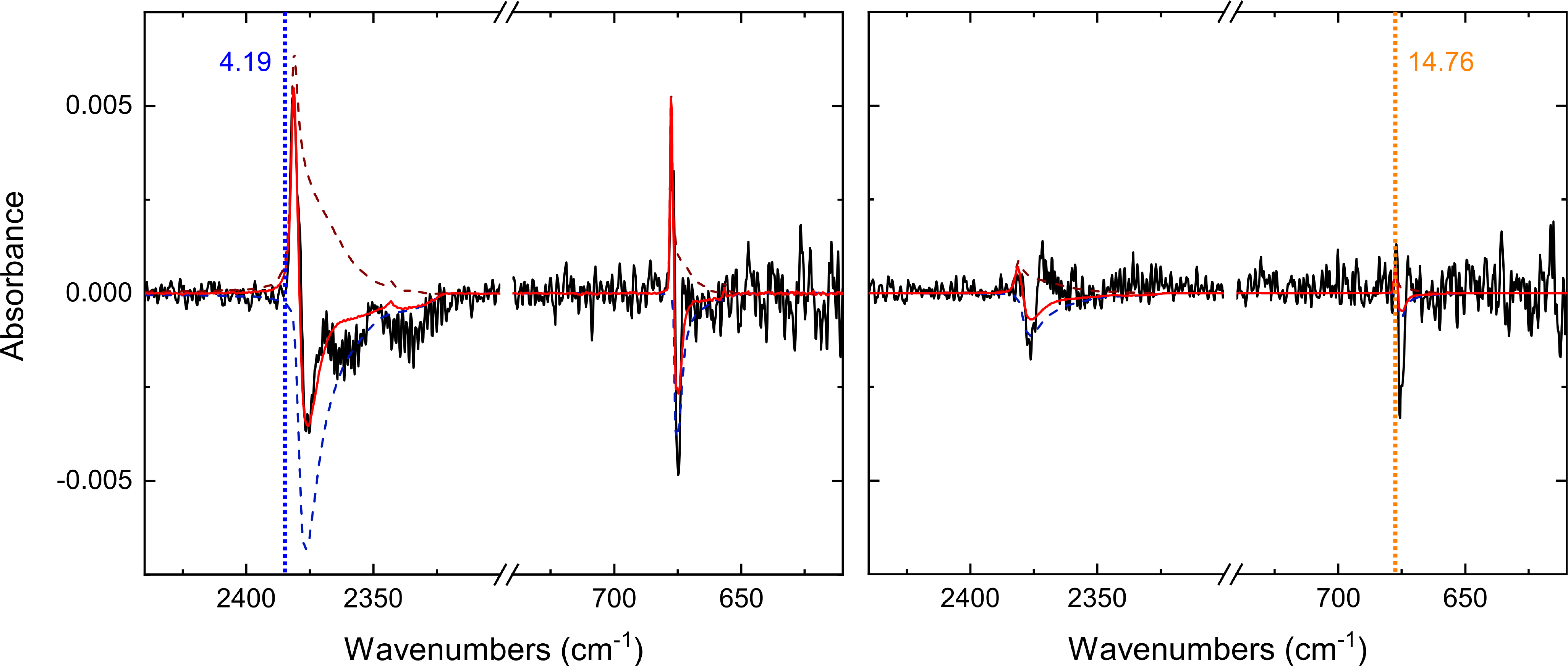}
	\caption{Spectral fits (red solid lines) of two distinct FEL irradiations at the stretching (left panel) and bending (right panel) modes of amorphous CO$_{2}$ ice at 20~K. FT-RAIR difference spectra acquired before and after FEL exposure are presented as solid black lines. Vertical blue and orange dotted lines indicate the selected FEL radiation frequencies. The dashed red line is a FT-RAIR spectrum acquired during the TPD of an unirradiated ice at 80~K, whereas the dashed blue line is a FT-RAIR spectrum acquired at 20~K. Both spectra are used as fit components.}
	\label{Fig4}
\end{figure*}

To understand whether FEL radiation effects cause a global heating of the ice or not, we have compared FEL irradiations at 4.19 and 14.76~$\mu$m (left and right panels of Figure~\ref{Fig4}, respectively) and fitted their profiles with selected FT-RAIR spectra of pure CO$_{2}$ ice deposited at 20~K and then heated during a TPD with a ramp of 2.5~K~min$^{-1}$ (see Table~\ref{Table1}). Each panel shows an independent fit with two spectral components, namely, a cold negative contribution at 20 K and a warm positive component at 80 K, which provided the best qualitative fits. The spectral component at 80~K shows that the ice in the control experiment has undergone restructuring during TPD toward a more crystalline-like form. The fact that a more crystalline ice (warm component) gives a positive contribution, while the amorphous ice (cold component) is a negative contribution to the fit suggests that there is a loss in amorphous ice because the ice structure reorganizes toward a more ordered configuration upon FEL exposure. Particularly, the fit at the $\nu_3$ mode of the ice irradiated at 4.19 $\mu$m shows a 2\% overall loss of column density of amorphous ice and a comparable increase in the column density of reorganized, crystalline-like, ice. When these numbers are scaled to the FEL exposed ice area on the substrate, morphological changes occur in $\sim$50\% of the irradiated molecules. Moreover, although the fit of the FEL irradiation at the $\nu_3$ mode (4.19~$\mu$m) seems to qualitatively well reproduce the stretching mode region, changes at the $\nu_2$ mode are not satisfactorily reproduced by the same fit. Furthermore, the FEL irradiation at the $\nu_2$ mode (14.76~$\mu$m) cannot be reproduced by any combination of all available FT-RAIR spectra acquired during the TPD control experiment as listed in Table~\ref{Table1}. This is because losses at the $\nu_2$ mode due to irradiation are always larger relative to the $\nu_3$ mode than in the case of global heating. Hence, changes in the ice induced by the FEL radiation are frequency dependent and cannot be simply explained by global thermal heating of the ice or global desorption.

Based on our previous results in similar studies, the most likely scenario is a selective vibrational excitation, that is a local increase in vibrational energy equivalent to local heating, and possibly minor desorption of specific molecules excited by the FEL radiation at the surface and/or within the bulk of the ice. FEL irradiation at the $\nu_3$ mode seems to induce mainly restructuring effects as shown by a combination of new positive and negative peaks in the difference spectra before and after FEL exposure, whereas irradiations at the $\nu_2$ mode seems to lead to an additional process, wherein a vibrational selectivity gives rise to the observation of only negative peaks \cite{Shirk:1990}. FEL-induced photodesorption could contribute to the observed reduction in the $\nu_2$-mode intensity upon irradiation. However, we should not exclude other possibilities such as hole-burning effects, particularly at the irradiated frequencies. Here, QMS data is key to directly detect FEL-induced photodesorption of CO$_{2}$. Unfortunately, although QMS data were successfully recorded during ice deposition and TPD experiments, during FEL irradiation, the acquisition speed of the instrument was not optimized for the detection of gas desorption plumes ejected from the ice in short bursts due to the pulsed nature of the FEL. Hence, during FEL irradiation, QMS results were inconclusive. The issue is beyond the scope of this paper and it will be addressed in future work.

\begin{figure*}
	\centering
	\includegraphics[width=\hsize]{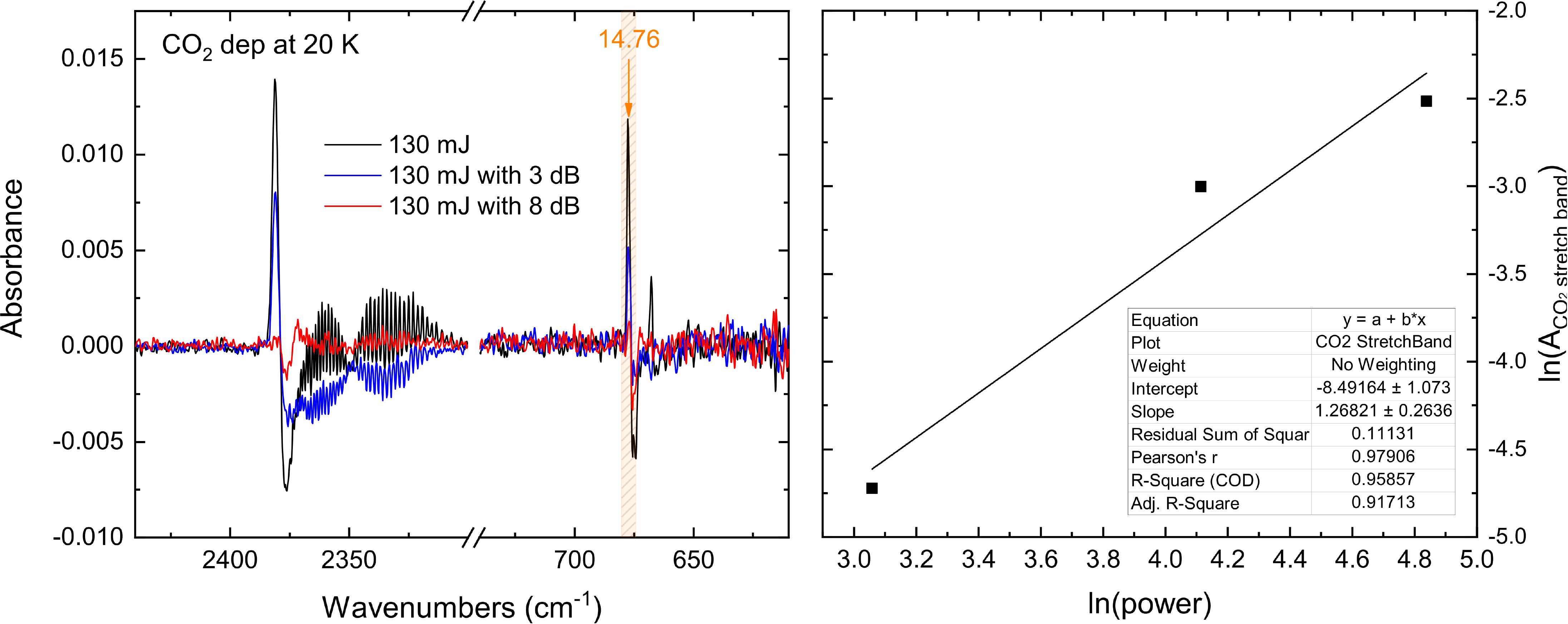}
	\caption{Power dependence of FEL radiation and spectral changes of amorphous CO$_{2}$ ice. Left panel: in solid black, blue, and red lines, FT-RAIR difference spectra acquired before and after FEL exposure of fresh spots of CO$_{2}$ ice with 0, 3, and 8 dB attenuation, respectively. Right panel: linear fit (solid black line) of the integrated areas (black full squares) of the three FT-RAIR spectra as a function of the FEL beam power. Both axes of the right panel are on a natural logarithmic scale.}
	\label{Fig5}
\end{figure*}

As for the case of pASW \cite{Noble:2020}, all morphological changes in the CO$_{2}$ ice are localized within the area exposed to the laser beam and do not cause a global eﬀect throughout the ice. This is verified by acquiring FT-RAIR spectra of the ice in regions in close proximity to irradiated areas and observing no FEL radiation effects. Moreover, to test whether in this case morphological changes in the ice are due to a single-photon process, we performed a FEL power dependence study on the irradiation of the CO$_{2}$ bending mode and measured changes at both the $\nu_3$ and $\nu_2$ modes. The left panel of Figure~\ref{Fig5} shows FT-RAIR difference spectra acquired before and after FEL exposure of three fresh ice spots at three different FEL beam power settings controlled by attenuating the beam power with 0, 3, and 8 dB, corresponding to powers of 130, 65, and 20 mJ as measured by a powermeter at the FELIX control station. The right panel of Figure~\ref{Fig5} shows that spectral changes measured by integrating the region around the $\nu_3$ mode are nearly linearly proportional to the increase in FEL power, in agreement with the hypothesis that changes in the ice are caused by a single-photon process. However, since local temperature at the irradiated ice spot cannot be measured, we cannot completely rule-out a contribution to the changes in ice morphology being caused by laser-induced global thermal heating. However, our FT-RAIR data show that full ice desorption, hole burning and ice melting effects do not occur at the FEL exposed spot of the deposited CO$_{2}$ layer as the spectral changes observed in our experiments reveal an IR photon-induced process that is selective and frequency dependent. This is also confirmed in preliminary experiments with more volatile species, such as CO and N$_{2}$, that are more sensitive to thermal changes in the ice than CO$_{2}$. Hence, our data indicate that thermal heating is a local process and vibrational energy dissipation likely affects only a few species in the ice per impinging IR photon. Future theoretical work on molecular dynamics simulations of our experiments will provide further insights on the IR photon-induced processing of space relevant ices.

\subsection{FEL irradiation of CO$_{2}$ ice deposited at 75~K}
\begin{figure*}
	\centering
	\includegraphics[width=\hsize]{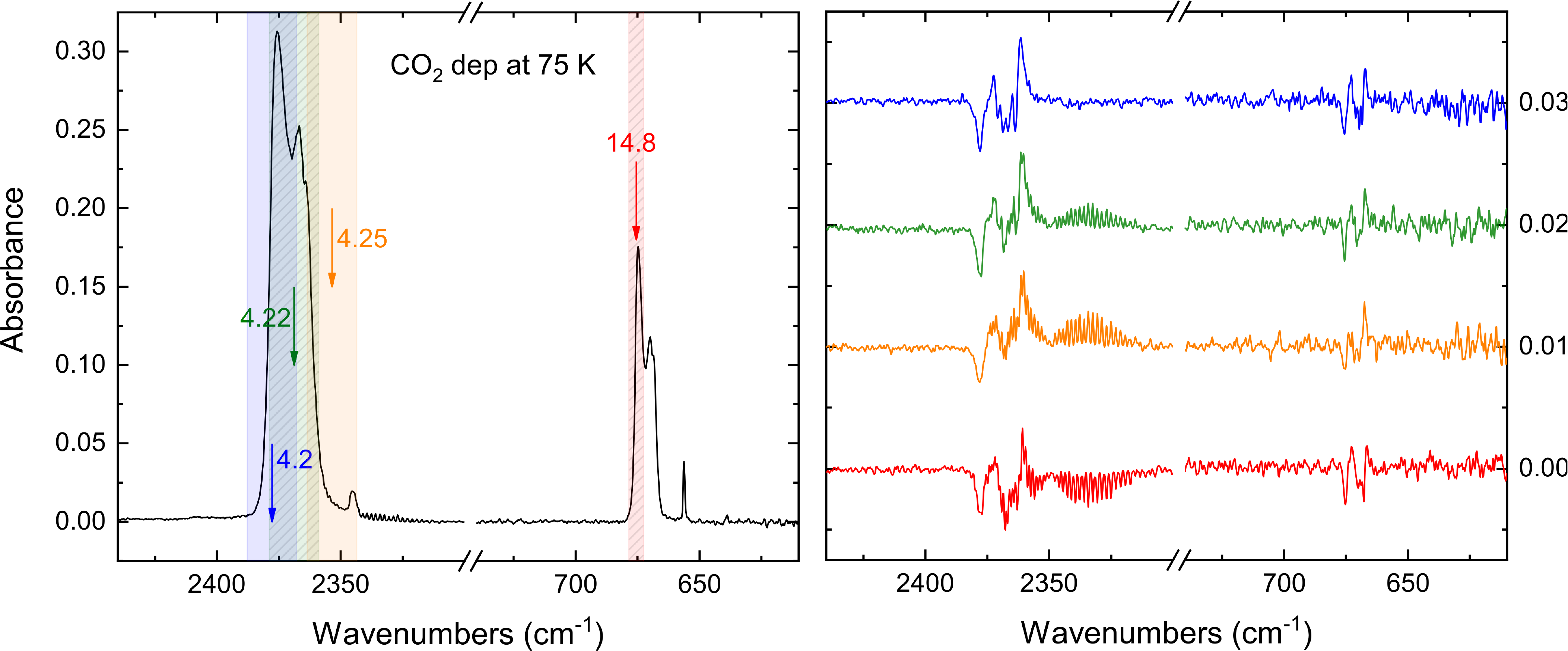}
	\caption{FEL irradiation of pure crystalline CO$_{2}$ ice deposited at 75~K and cooled to 20~K. Left panel shows a FT-RAIR spectrum of pure CO$_{2}$ ice after deposition in the $\nu_3$ and $\nu_2$ spectral ranges. Colored arrows and strips indicate the frequencies and FWHM of FELIX-2 used during irradiation, respectively. Right panel shows FT-RAIR difference spectra before and after FEL exposure. Spectra and FEL irradiation frequencies are color-coded. Difference spectra are offset for clarity.}
	\label{Fig6}
\end{figure*}

The left panel of Figure~\ref{Fig6} shows a FT-RAIR spectrum of pure CO$_{2}$ ice deposited at 75~K and then cooled to 20~K in both the $\nu_3$ and $\nu_2$ spectral ranges. Colored arrows indicate the frequencies at which the ice has been exposed to FELIX-2 radiation for five minutes of continuous exposure at a time and always on fresh ice spots. Vertical color-coded strips indicate the spectral FWHM of the FELIX beam at each selected wavelength. As in Figure~\ref{Fig3}, the color of the arrows in the left panel of Figure~\ref{Fig6} corresponds to the color of the FT-RAIR difference spectra acquired before and after FEL exposure as shown in the right panel of the same figure. Spectral changes highlighted in the difference spectra of the right panel of Figure~\ref{Fig6} are all qualitatively comparable regardless of whether the ice is irradiated at the $\nu_3$  or $\nu_2$  mode. This suggests that reorganization of the ice is unlikely to occur, which was expected since the ice is already crystalline due to its deposition temperature of 75~K.

\begin{figure}
	\centering
	\includegraphics[width=\hsize]{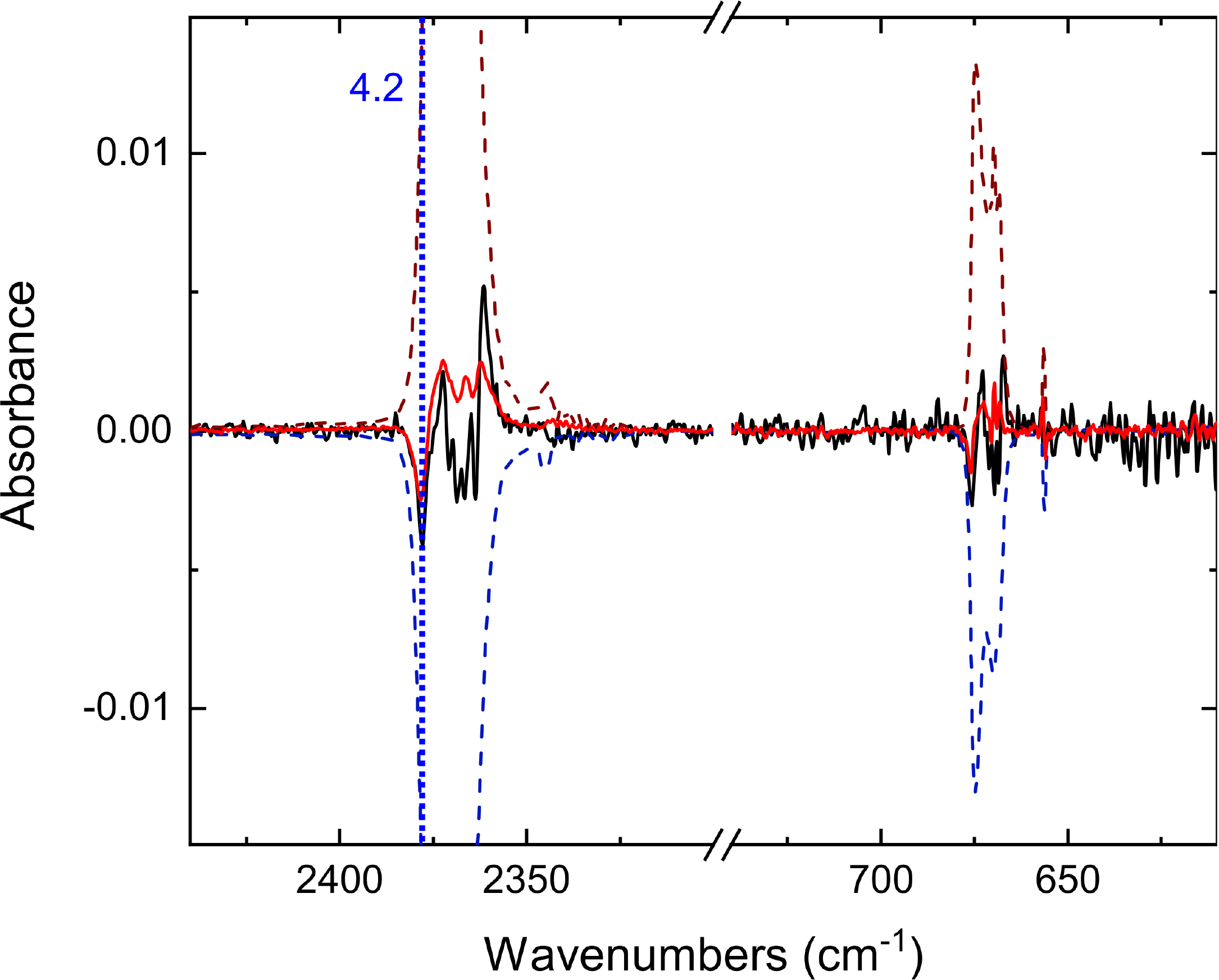}
	\caption{Spectral fit (red solid line) of a FEL irradiation at the stretching mode of crystalline CO$_{2}$ ice deposited at 75~K and cooled to 20~K. A FT-RAIR difference spectrum acquired before and after FEL exposure is reported as a solid black line. Vertical blue dotted line indicates the selected FEL radiation frequency. Dashed red line is a FT-RAIR spectrum acquired after deposition of the ice at 75~K, whereas the dashed blue line is a FT-RAIR spectrum acquired after the ice is cooled to 20~K. Both spectra are used as fit components.
	}
	\label{Fig7}
\end{figure}

Figure~\ref{Fig7} shows the FT-RAIR difference spectrum of the pure CO$_{2}$ ice deposited at 75~K and then cooled to 20~K before and after being exposed to FEL radiation at 4.2~$\mu$m. The profile of the difference spectrum is fitted with two spectral components, namely, a cold negative contribution of the same ice before FEL irradiation at 20~K and a warm positive contribution of the ice as deposited at 75~K. Unlike the case of amorphous CO$_{2}$ ice, which is a metastable material, in the case of crystalline CO$_{2}$ ice, changes in the spectral profile at different temperatures are reversible and can be obtained during multiple temperature cycles. As the cryostat cools the ice sample after deposition of a crystalline ice, IR features generally become sharper and more intense at lower temperatures (10-20~K) relative to the same features at higher temperatures (\emph{e.g.}, 75~K in the case of CO$_{2}$) because cooling the sample will reduce the thermal population of hot-band states of the vibrations. Such reversible process is also seen in the THz spectral range \cite{Allodi:2014, Ioppolo:2014}. The lowering of the temperature reduces the thermal agitation of the lattice and consequently the phonon population in the THz range, which lowers the number of nearly equivalent classes of oscillators. This allows for a better minimization of the local potential energy.

It is worth noting that the two crystalline spectral components acquired at different temperatures of the same deposited ice prior to FEL exposure can partially fit the changes seen in the crystalline ice after FEL irradiation. However, as opposed to the case of amorphous ice seen in Figure~\ref{Fig3}, spectral changes upon FEL exposure of crystalline CO$_{2}$ ice seem to be less dependent on the selected absorption mode. A closer look at the fit of Figure~\ref{Fig7} suggests that some negative peaks in both the $\nu_3$ and $\nu_2$ modes are never reproduced by the chosen fit components, indicating that some other processes such as desorption may be occurring in the ice. Here, the loss in the $\nu_3$ mode is larger than that in the $\nu_2$ band, as expected if the ice desorbed. FEL photoinduced desorption was already observed in the case of crystalline H$_{2}$O ice \cite{Noble:2020, Focsa:2003}. As mentioned before, QMS data are not conclusive concerning the detection of IR photodesorption. Nevertheless, indirect evidence is shown in the FT-RAIR difference spectra for both amorphous and crystalline CO$_{2}$ ice. Dedicated studies of infrared-induced photodesorption of solid molecules will be the focus of future research campaigns at HFML-FELIX Laboratory.

The difference in behavior of the pure CO$_{2}$ ices -- that deposited at 20~K and that at 75~K and then cooled to 20~K -- upon FEL irradiation of selected absorption vibrational modes in the MIR is new, independent evidence that CO$_{2}$ ice deposited at 20~K under our experimental conditions is an amorphous material, because it undergoes reorganization and minor desorption, whereas crystalline CO$_{2}$ ice does not show any clear sign of restructuring but mainly desorption. McGuire $et~al.$~\cite{McGuire:2016} conducted a thorough investigation of the THz features of amorphous and crystalline CO$_{2}$ ice, finding unambiguous evidence that the only two features of CO$_{2}$ ice in the far-IR at 3.3 and 2.1 THz are due solely to crystalline CO$_{2}$, and that amorphous CO$_{2}$ is featureless. Our work confirms that a background slow deposition of CO$_{2}$ molecules under UHV conditions leads to the formation of amorphous CO$_{2}$ ice. However, we cannot rule-out that solid CO$_{2}$ samples grown under different conditions at low temperature, for instance, through direct deposition (\emph{i.e.}, normal to the substrate face) and at higher deposition rates, may lead to the formation of the polycrystalline form of ice already reported in literature \cite{Escribano:2013, Gerakines:2015, Baratta:2017, Tsuge:2020, Kouchi:2021}. A further systematic investigation of different deposition conditions and ice thicknesses is needed to clarify this point.

\section{Conclusions}

This study focuses on the selective MIR irradiation of pure CO$_{2}$ ices deposited at 20~K and at 75~K then cooled to 20~K by means of the intense, tunable radiation of the FELIX-2 free-electron laser at the HFML-FELIX Laboratory. Experiments are carried out using the LISA end station, where ice samples are grown and monitored $in~situ$ by Fourier Transform Reﬂection-Absorption Infrared spectroscopy in the MIR range and by quadrupole mass spectrometry. We have shown that amorphous CO$_{2}$ ice goes through a restructuring process reorienting some of its molecules into a more ordered structure upon resonant absorption of IR radiation. The exposure of crystalline CO$_{2}$ ice to FEL radiation induces mostly desorption, while frequency-dependent restructuring of the ice material is not clearly observed. This work adds to the current debate on the nature of CO$_{2}$ ice in laboratory experiments and in space.

\section*{Supplementary Material}
\subsection*{Details of the Experimental Setup}
\begin{figure}
	\centering
	\includegraphics[width=\hsize]{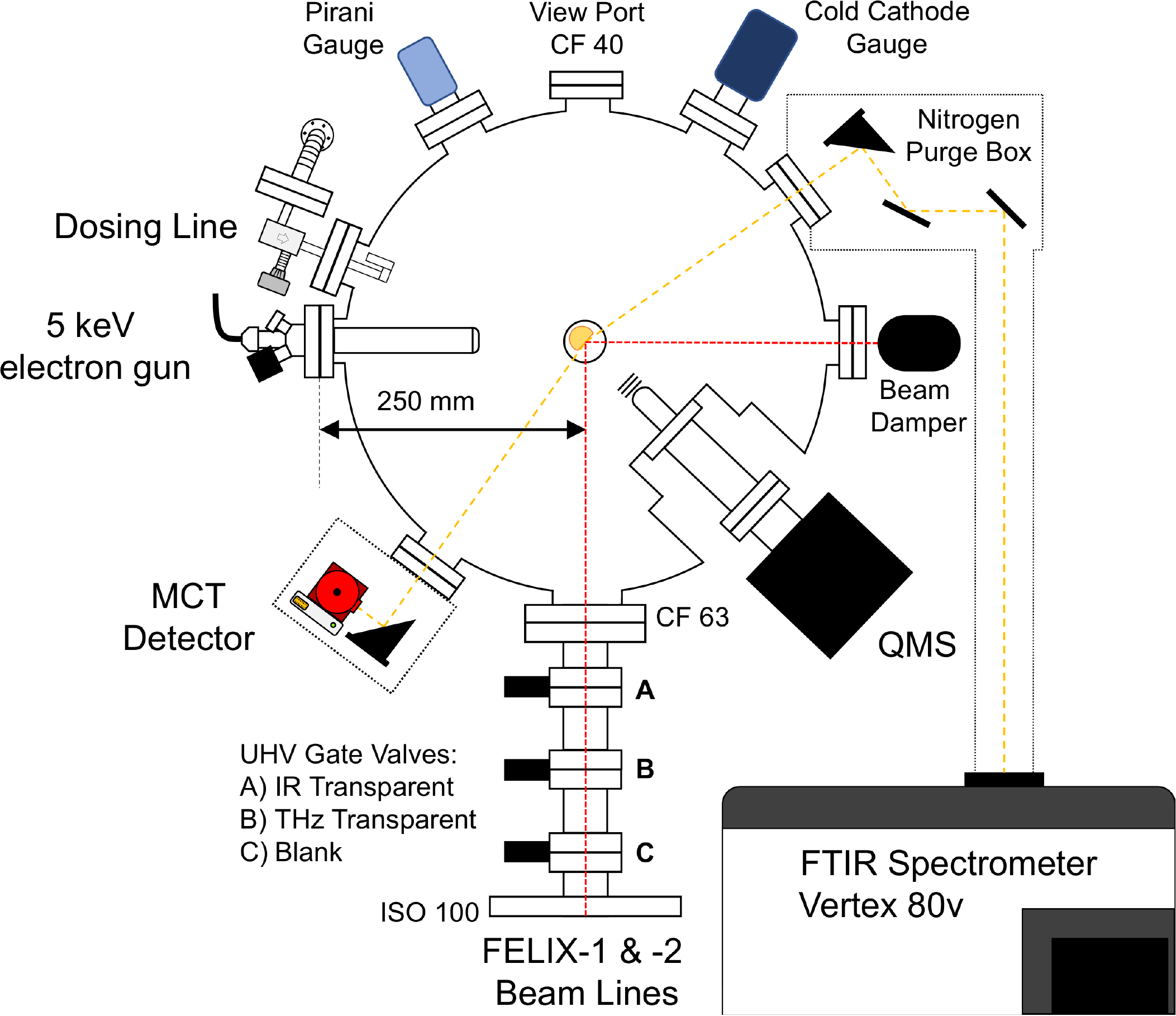}
	\caption{Schematic top view of the Laboratory Ice Surface Astrophysics (LISA) end station at the HFML-FELIX Laboratory in the Netherlands.
	}
    \label{Fig1}
\end{figure}

Experiments are carried out using the Laboratory Ice Surface Astrophysics (LISA) ultrahigh vacuum (UHV) end station at the HFML-FELIX Laboratory, Radboud University in the Netherlands. The experimental setup has been rebuilt since its original description in Noble $et~al.$~\cite{Noble:2020}. In the present paper, we present a detailed description of the experimental apparatus. The LISA system has been designed and optimized to perform selective IR/THz irradiation of solid phase molecules when coupled to the tunable, high-power, and short-pulsed radiation from the FELs FELIX-1 ($\sim$30-150~$\mu$m) and FELIX-2 ($\sim$3-45~$\mu$m). Figure~\ref{Fig1} shows a schematic of the experimental configuration used at the time of the present work. The UHV system ($\sim$10$^{-10}$~mbar), substantially larger than in previous iterations, comprises a $\sim$500~mm diameter cylindrical main chamber with upper and lower DN100CF and DN200CF flanges, respectively, as well as another nine DN40CF ports and a DN63CF connection on its sides. At its base, the chamber is connected to a DN200CF turbo molecular pump (TURBOVAC MAG $W~1500$, Oerlikon Leybold Vacuum) through a DN200CF all-metal gate valve. The turbo pump is then connected to a foreline pump (D~40~BCS, Oerlikon Leybold Vacuum) with an ultimate pressure of 10$^{-4}$~mbar by means of a DN40KF manual ball valve and an activated alumina oil trap.

On top of the DN100CF flange of the main chamber there is an $xyz$-translator (LewVac) with a $z$-stroke of 50.8~mm and $xy$-fine control of $\pm12.7$~mm connected to a two stage differentially-pumped rotary platform DN100CF with half nipple base flange (LewVac), and a closed-cycle helium CH-204SB cryostat head with a HC-4E water-cooled compressor (Sumitomo). Two T-Station 85H pumps (Edwards Vacuum) are used to differentially pump the rotary platform and to evacuate the DN100ISO-K end connections between LISA and the FEL beam line, which also include a DN100ISO-K/DN40CF adapter, a series of three DN40CF gate valves, and a DN40CF/DN63CF adapter (see Fig.~\ref{Fig1}). In particular, two of the three manual UHV DN40CF gate valves mounted in series in the FEL beam path before the main chamber have CsI (Crystran) and TPX (Tydex) windows, respectively transparent in the IR and THz spectral ranges, to allow for a faster change between FELIX-1 and -2 during a single beam time shift (8 hours) with no need to open the vacuum line. The blank DN40CF gate valve is used to block the FEL beam when ices are not being irradiated. Moreover, the use of these valves between the chamber and the FEL beam line allows LISA to remain a truly UHV system at all times, with a negligible water contamination in the main chamber, because the valves isolate it from the high vacuum pressure of the DN100ISO-K FEL beam line ($\sim$10$^{-6}$~mbar).

The cryostat head is in thermal contact with a custom-made 258~mm long oxygen-free high thermal conductivity (OFHC) copper circular rod (GoodFellow) with a diameter of 38~mm. At one end of the rod, a single optically flat gold-coated 38$\times$19$\times$57~mm (l$\times$w$\times$h) face is used as substrate to ensure maximal FEL beam reflection in the IR/THz spectral range at the center of the main chamber. Connected to the first stage of the cryostat head, an OFHC copper radiation shield surrounds the copper rod leaving only the substrate face uncovered. With the cryostat on, the temperature of the radiation shield is kept constant at $\sim$77~K. In this experimental configuration, the selected substrate face exposed to the FEL beam can be moved in height by means of the $z$-translator to allow for the irradiation of the gold-coated surface at different spots on the substrate face. The $z$-translation stage is also used to remove the substrate from the line of sight of the FEL beam such that the alignment of the Helium-Neon (HeNe) laser of FELIX-1 and -2 is possible at any time through a ZnSe viewport (Hositrad) at the back of the chamber, opposite the entrance port of the FEL source.

The substrate temperature is controlled in the range 15-300~K by means of a model 336 cryogenic temperature controller (LakeShore) that reads temperature through an uncalibrated DT-670B-SD silicon diode (LakeShore) mounted with a CO adapter on the upper backside of the gold-coated flat substrate and regulates a $25~\Omega/100$~W cartridge heater (HTR-25-100) mounted inside the copper rod, at the mid-height of the flat substrate. Although not currently in use, a second silicon diode can be mounted on the lower backside of the substrate, in a position symmetric to the first diode and the center of the flat substrate. Pressure in the main chamber is monitored by means of a wide range (5$\times$10$^{-3}$ to 1$\times$10$^{-11}$~mbar) cold cathode gauge DN40CF (IKR 070, Pfeiffer Vacuum) coupled to a TPG 300 Total pressure system with a CP 300 T11 Pirani/Cold cathode measurement board (Pfeiffer Vacuum). Active Pirani vacuum gauges (TPR 018, Pfeiffer Vacuum) are connected at various points of the system, for instance in the main chamber, between the turbo pump and the foreline pump, and in the dosing line.

A DN16CF tubing dosing line is used to prepare pure gases and gas mixtures prior to their dosing inside the main chamber. The dosing line is first roughly pumped by means of a dry scroll pump (Edwards Vacuum) and then base pressure better than 10$^{-4}$~mbar is achieved using a T-Station 85H (Edwards Vacuum). Apart from a Pirani gauge that measures the base pressure in the dosing line, two active capacitance transmitters (0.001-10 and 0.1-1000~mbar, models CCR~363 and CCR~361, respectively, Pfeiffer Vacuum) coupled to a CenterThree display and control unit (Pfeiffer Vacuum) are used to measure gases mass-independently during gas mixture preparation, ensuring reproducibility of the desired ice mixture ratios. Gases are introduced into the main chamber through an all-metal leak valve (LewVac) connected to a custom-made double-sided DN40CF flange (LewVac) with a 6~mm central tube facing the walls of the chamber to allow for background deposition of gases onto the substrate. In this way, molecules are deposited at random angles with respect to the substrate surface to better simulate deposition in space on interstellar dust grains. A background deposition also allows for a more uniform ice deposition all over the substrate surface. Hence, once a selected ice is deposited, FEL irradiation can be carried out at multiple unirradiated spots of the same deposited ice, maximizing the use of a beam time shift and allowing for more systematic and reproducible studies.

Four DN40CF flanges of the main chamber have ZnSe viewports (Hositrad) mounted to perform infrared spectroscopy and measure FELIX-2 beam power after the sample. When the FEL beam power is not measured, a beam damper is placed at the exit of the ZnSe viewport to block any radiation exiting the chamber, as shown in Figure~\ref{Fig1}. An external liquid nitrogen cooled mercury cadmium telluride (MCT) detector is used to acquire mid-IR (MIR) data (4000-600~cm$^{-1}$, 2.5-16.6 $\mu$m) by means of a Fourier transform infrared spectrometer (FTIR) in reflection-absorption infrared mode (RAIR) and at an angle from the gold surface to the normal of 13$^\circ$ (VERTEX 80v, Bruker). Flat and off-axis parabolic gold mirrors are used to steer and focus, respectively, the FTIR beam onto the substrate and the MCT (see Fig.~\ref{Fig1}). While the spectrometer is pumped with a dry scroll pump (Edwards Vacuum), external optics and the MCT detector are in purge boxes filled with pure nitrogen gas. Figure~\ref{Fig1} shows a residual gas analyzer quadrupole mass spectrometer (QMS, MKS Instruments) facing the gold flat substrate and at a distance of $\sim$55~mm from its center to detect desorbing species during FEL irradiation of the ice and temperature programmed desorption (TPD) experiments. Alignment of FEL and FTIR beams is done by manipulating the $xyz$-translator to measure changes in the FTIR signal and in the intensity of the FEL HeNe, and by steering a series of gold mirrors in the two lines to maximize their overlap at the substrate central spot. The FEL and FTIR beams as well as the QMS are all aligned to the center of the chamber and are all in the same horizontal plane. We measured the FELIX beam spot size at the substrate surface to be approximately 2~mm in height (diameter), while the FTIR beam height (diameter) is $\sim$3~mm, with both beams elongated in width due to their incident angles on the substrate, 45$^\circ$ and 13$^\circ$, respectively. Assuming ellipsoidal cross sections of the two beams at the substrate face, the FEL beam area covers only $\sim$14\% of the area of the FTIR beam. The current FEL and FTIR beam spot sizes allow for the irradiation and investigation of at least seven fresh, unirradiated spots per single ice deposition. Although not used in this work, LISA also includes a 50~eV to 5~keV electron gun (EFG-7/EGPS-2017, Kimball Physics) with a FC-7 manual Faraday cup at its end to induce chemical reactions within deposited ices.

\section*{CRediT authorship contribution statement}
S. Ioppolo initiated and managed the project (FELIX-2020-01-51) at HFML-FELIX Laboratory. He wrote the manuscript with assistance from H.M. Cuppen, J.A. Noble, and B. Redlich. S. Ioppolo and J.A. Noble performed all laboratory experiments. A. Traspas~Mui\~{n}a assisted during assembling and testing of the LISA end station and performed control experiments for the project. S. Coussan performed some preliminary experiments. All authors contributed to data interpretation and commented on the paper.

\section*{Declaration of Competing Interest}
The authors declare that they have no known competing financial interests or personal relationships that could have appeared to influence the work reported in this paper.

\section*{Acknowledgements}
The authors thank the HFML-FELIX Laboratory team for their experimental assistance and scientific support. The LISA end station is designed, constructed, and managed at the HFML-FELIX Laboratory by the group of S. Ioppolo. This work was supported by the Royal Society University Research Fellowships Renewals 2019 (URF/R/191018); the Royal Society University Research Fellowship (UF130409); the Royal Society Research Fellow Enhancement Award (RGF/EA/180306); and the Royal Society Research Grant (RSG/R1/180418). Travel support was granted by the UK Engineering and Physical Sciences Research Council (UK EPSRC Grant EP/R007926/1 - FLUENCE: Felix Light for the UK: Exploiting Novel Characteristics and Expertise).





\bibliographystyle{elsarticle-num}
\bibliography{mybibfile}

\vskip3pt

\end{document}